\documentclass[fleqn,twoside]{article}
\usepackage[headings]{espcrc2}

\readRCS
$Id: espcrc2.tex,v 1.2 2004/02/24 11:22:11 spepping Exp $
\usepackage{graphicx}
\usepackage[figuresright]{rotating}

\usepackage{amsmath}
\usepackage{amssymb}

\title{Combined HERA Deep Inelastic Scattering Data and NLO QCD Fits}

\author{C. Gwenlan\address[MCSD]{Department of Physics and Astronomy, 
        University of Oxford, Oxford. OX1 3RH. United Kingdom. \\
        }, on behalf of the HERA Combined Structure Functions Working Group}

\runauthor{C. Gwenlan}

\begin{document}

\begin{abstract}
Previously published data on inclusive neutral and charged current 
$e^+p$ and $e^-p$ deep inelastic scattering from HERA have been combined.  
The new, model-independent method of combining the measured cross sections 
takes full account of correlated systematics in a coherent way, 
leading to significantly reduced uncertainties in the combined measurement 
across the $(x,Q^2)$ plane. The combined cross section data have 
been used as the sole input for a new next-to-leading order QCD fit to 
extract the parton distribution functions (PDFs) of the proton. The PDFs extracted 
have greatly reduced experimental uncertainties,    
compared to separate QCD analyses on data from H1 and ZEUS. 
Model uncertainties, including those arising from the parameterisation 
dependence, have also been carefully considered. The resulting HERA PDFs have impressive precision. 
\vspace{1pc}
\end{abstract}

\maketitle

\section{Introduction}
The deep inelastic lepton-nucleon scattering (DIS) process, 
in which a point-like lepton probes a single, initial state hadron, 
is an ideal environment in which to study quantum chromodynamics 
(QCD), and constrain the parton distribution functions (PDFs) of the proton.

The HERA collider ceased running in June 2007, following 
15 years of successful operation. During HERA I running (1992-2000), 
the H1 and ZEUS experiments each collected approximately 
$100$ ${\rm pb}^{-1}$ of $e^+p$ and  $\sim 15$ ${\rm pb}^{-1}$ of $e^-p$ 
data. This has allowed precision measurements of inclusive neutral (NC) 
and charged (CC) current DIS cross sections, which have 
already proved fundamental to the rapid development in 
understanding of QCD and the structure of the proton.

In these proceedings, a joint H1 and ZEUS analysis is presented, in which previously 
published NC and CC inclusive DIS cross section measurements from the 
two collider experiments are combined. 
The combination method~\cite{ref:dis05} uses an iterative $\chi^2$ minimisation, which 
carefully takes into account the correlations within the data that 
result from different sources of uncertainty.
The key assumption 
is that the measurements from H1 and ZEUS represent a common truth. 
Thus, forcing them to agree results in a strong 
constraint which cross-calibrates the measurements, resulting 
in significantly reduced overall uncertainties.
The combined HERA data have subsequently been used as the sole input to a  
next-to-leading (NLO) QCD fit to determine the proton PDFs.   
This analysis is also discussed in the present contribution.  

The results on the combined HERA data, as presented here, 
were first released to the LP Conference in 2007~\cite{ref:combined:h1andzeus}, 
and the NLO QCD analysis was first presented at DIS 2008~\cite{ref:qcdanalysis:dis08}. 
Both analyses should be seen as part of a mid-term strategy. 
They will be followed by future data combinations, including  
even more accurate data from both HERA I 
and HERA II, and by further QCD analyses to extract the proton PDFs.

\section{HERA physics and kinematics}
Lepton-proton DIS can proceed either 
via the neutral current (NC) interaction (through the exchange of 
a $\gamma^*$ or ${Z}^0$), or via the charged current 
(CC) interaction (through the exchange of a ${W}^{\pm}$). 
The kinematics of lepton-proton DIS are described 
in terms of the Bjorken scaling variable, $x$, the negative invariant 
mass squared of the exchanged vector boson, $Q^2$, and the 
fraction of energy transferred from the lepton to the proton system 
(in the rest frame of the proton), $y$. 
The three quantities are related by $Q^2=s \cdot x \cdot y$, where $s$ is 
the centre-of-mass energy squared.

At leading order (LO) in the electroweak interaction, 
the reduced cross section, $\sigma_r(x,Q^2)$, for 
the $e^\pm p$ NC DIS process can be expressed in terms 
of proton structure functions,
\begin{eqnarray}
{\sigma}_{r,{\rm NC}}^{e^\pm p} &=&  {F}_2 \mp \frac{Y_-}{Y_+}x{F}_3-\frac{y^2}{Y_+}{F}_{\rm L} \nonumber \\
&=& \frac{xQ^4}{2\pi \alpha^2}\frac{1}{Y_+}\frac{{\rm d^2}\sigma_{\rm NC}^{e^\pm p}}{{\rm d}x{\rm d}Q^2},  \nonumber 
\label{eqn:nc}
\vspace{-0.2cm}
\end{eqnarray}
where $\alpha$ is the fine structure constant and $Y_{\pm}=1\pm(1-y)^2$ 
describes the helicity dependence of the electroweak interaction. 
The structure functions, which depend on $(x,Q^2)$, 
are directly related to PDFs of the proton, and 
their $Q^2$ dependence is predicted by perturbative QCD. 
In particular, ${F}_2$ and $x{F}_3$ depend directly on 
the quark distributions. For $Q^2 < 1000$ ${\rm GeV}^2$, ${F}_2$ 
dominates the $ep$ scattering cross section and for $x < 10^{-2}$, ${F}_2$ itself is 
dominated by sea quarks while the $Q^2$ dependence is 
driven by gluon radiation. ${F}_2$ has been measured over 4 orders of 
magnitude in $(x,Q^2)$, to a precision of $\sim 2-3\%$ at HERA I. 
Therefore, HERA data have already provided vital information on the sea-quarks and gluon at low $x$. 
At high $Q^2 \gtrsim M_{Z}^2$, 
the contribution from $x{F}_3$ becomes increasingly significant and gives 
information on the valence quarks. 

At LO, the CC cross sections are given by,
\begin{eqnarray}
\frac{{\rm d}^2\sigma_{\rm CC}^{e^{+}p}}{{\rm d}x{\rm d}Q^2} 
= \frac{G_F^2M_W^4}{2\pi(Q^2+M_W^2)^2}  \left [ \bar{u}+\bar{c}+(1-y)^2(d+s) \right ] ~~ \nonumber \\ 
\frac{{\rm d}^2\sigma_{\rm CC}^{e^{-}p}}{{\rm d}x{\rm d}Q^2} 
= \frac{G_F^2M_W^4}{2\pi(Q^2+M_W^2)^2}  \left [ {u}+{c}+(1-y)^2(\bar{d}+\bar{s}) \right ] ~~ \nonumber
\vspace{-0.2cm}
\label{eqn:cc}
\end{eqnarray}
so that a measurement of the $e^+p$ and $e^-p$ cross sections 
provides information on the $d$- and $u$-valence 
quarks, respectively, thereby allowing the separation of flavour. 

\section{Combination of HERA DIS data}
\label{sec:combineddata}
In this section, the joint H1 and ZEUS combined data analysis is described. 
The goal of the study is to obtain DIS cross sections of best possible accuracy 
in order produce precise extractions of the proton PDFs. A new NLO QCD 
analysis, in which the results of the combination are used as the sole input, 
is described in Sec.~\ref{sec:qcdanalysis}.

\subsection{Combination Method}
\label{sec:combinationmethod}
The averaging procedure uses the Lagrange Multiplier (or Hessian) method~\cite{ref:hessian}. 
The only theoretical input to the combination 
is that there is a true value of the cross section for each process, at each $(x,Q^2)$ 
value~\cite{ref:combined:h1andzeus,ref:combined:dis08}. 
The correlated systematic uncertainties are floated coherently such that 
each experiment calibrates the other one. 
This results in a significant reduction of the correlated systematic 
uncertainty over much of the kinematic plane. 

In the combination procedure, the following probability distribution of a 
measurement quantity, $M$, represented as a $\chi^2$ function, is minimised: 
\begin{eqnarray}
\label{eqn:chi2}
\chi^2_{exp}(M^{i,{\rm true}},\alpha_j) =   ~~~~~~~~~~~~~~~~~~~~~~~~~~~~~~~~~~~~~ \nonumber \\
\sum_{i}\frac{\left [ M^{i,{\rm true}} - \left ( M^i + \sum_j \frac{\partial M^i}{\partial \alpha_j}\alpha_j  \right ) \right ]^2}{\delta_i^2} + \sum_j \frac{\alpha_j^2}{\delta^2_{\alpha_j}}.
\end{eqnarray}
Here, $M^i$ are the measured central values, and $\delta_i$ are the statistical 
and uncorrelated systematic uncertainties of the quantity $M$. The $M^{i,{\rm true}}$ 
are their true values; $\alpha_j$ are parameters for the $j$ sources of systematic 
uncertainty and $\partial M^i/\partial\alpha_j$ denotes the sensitivity of point 
$i$ to source $j$. For the cross section measurements, the index $i$ labels a 
particular measurement at a given $(x,Q^2)$. Equation~\ref{eqn:chi2} represents the 
correlated probability distribution function for the quantity $M^{i,{\rm true}}$ and 
for the systematic uncertainties $\alpha_j$.

The $\chi^2$ defined in Eq.~\ref{eqn:chi2} has, by construction, a minimum $\chi^2=0$ 
for $M^{i,{\rm true}}=M^i$ and $\alpha_j=0$. The total uncertainty for $M^{i,{\rm true}}$ 
determined from the formal minimisation of Eq.~\ref{eqn:chi2} is equal to 
the quadratic sum of the statistical and systematic uncertainties. 
The covariance matrix cov($M^{i,{\rm true}},M^{k,{\rm true}}$) quantifies the correlation 
between experimental points.

In the analysis of more than one data set, a total $\chi^2$ function, $\chi^2_{\rm tot}$, 
is constructed from the sum of the $\chi^2$ functions for each data set. The $\chi^2_{\rm tot}$ 
function can be minimised with respect to $M^{i,{\rm true}}$ and $\alpha_j$: 
this minimisation corresponds to a generalisation of the averaging procedure 
which takes account of correlations between different data sets. 

\begin{center}
\begin{table*}[Htp]
\begin{tabular}{|lr|lr|lr|c|c|c|}
\hline
\multicolumn{2}{|c|}{\small data set} &\multicolumn{2}{c|} {\small $x$ range} &\multicolumn{2}{c|} {\small $Q^2$ range (${\rm GeV}^2$)}  & {\small $\mathcal{L}$ (${\rm pb}^{-1}$)} & {\small comment}  & {\small ref.} \\ \hline
{\small H1 NC min. bias} & 97     & 0.00008  &0.02 & 3.5  &    12 & 1.8  & $e^+ p$ $\sqrt{s}=301$ GeV  & \cite{epj:c21:33} \\
{\small H1 NC low $Q^2$} & 96-97  & 0.000161 &0.20 & 12   &   150 & 17.9 & $e^+ p$ $\sqrt{s}=301$ GeV  & \cite{epj:c21:33} \\
{\small H1 NC }          & 94-97  & 0.0032   &0.65 & 150  & 30000 & 35.6 & $e^+ p$ $\sqrt{s}=301$ GeV  & \cite{epj:c13:609} \\
{\small H1 CC }          & 94-97  & 0.013    &0.40 & 300  & 15000 & 35.6 & $e^+ p$ $\sqrt{s}=301$ GeV  & \cite{epj:c13:609} \\
{\small H1 NC }          & 98-99  & 0.0032   &0.65 & 150  & 30000 & 16.4 & $e^- p$ $\sqrt{s}=319$ GeV  & \cite{epj:c19:269} \\
{\small H1 CC }          & 98-99  & 0.013    &0.40 & 300  & 15000 & 16.4 & $e^- p$ $\sqrt{s}=319$ GeV  & \cite{epj:c19:269}  \\
{\small H1 NC }          & 99-00  & 0.00131  &0.65 & 100  & 30000 & 65.2 & $e^+ p$ $\sqrt{s}=319$ GeV  & \cite{epj:c30:1}  \\
{\small H1 CC }          & 99-00  & 0.013    &0.40 & 300  & 15000 & 65.2 & $e^+ p$ $\sqrt{s}=319$ GeV  & \cite{epj:c30:1} \\  \hline
{\small ZEUS NC }        & 96-97  & 0.00006  &0.65 & 2.7  & 30000 & 30.0 & $e^+ p$ $\sqrt{s}=301$ GeV  & \cite{epj:c21:443} \\
{\small ZEUS CC }        & 94-97  & 0.015    &0.42 & 280  & 17000 & 47.7 & $e^+ p$ $\sqrt{s}=301$ GeV  & \cite{epj:c12:411} \\
{\small ZEUS NC }        & 98-99  & 0.005    &0.65 & 200  & 30000 & 15.9 & $e^- p$ $\sqrt{s}=319$ GeV  & \cite{epj:c28:175} \\
{\small ZEUS CC }        & 98-99  & 0.015    &0.42 & 280  & 30000 & 16.4 & $e^- p$ $\sqrt{s}=319$ GeV  & \cite{pl:b539:197}  \\
{\small ZEUS NC }        & 99-00  & 0.005    &0.65 & 200  & 30000 & 63.2 & $e^+ p$ $\sqrt{s}=319$ GeV  & \cite{pr:d70:052001} \\
{\small ZEUS CC }        & 99-00  & 0.008    &0.42 & 280  & 17000 & 60.9 & $e^+ p$ $\sqrt{s}=319$ GeV  & \cite{epj:c32:1} \\  \hline
\end{tabular}
\vspace{0.5cm}
\caption{The H1 and ZEUS datasets used in the combination procedure. 
The integrated luminosity ($\mathcal{L}$) and the kinematic range in $(x,Q^2)$ are given. 
Note that a re-analysis of the H1 luminosity measurement of the special 
minimum bias run in 1997 has lead to a change in integrated luminosity. This has  
resulted in an upward shift of the H1 minimum bias data by $3.4\%$, 
which is taken into account in the combination.}
\label{tab:ncccdatasets}
\end{table*}
\vspace{-0.8cm}
\end{center}
The $\chi^2$ function of Eq.~\ref{eqn:chi2} is most suitable for absolute or 
{\it additive} uncertainties, i.e. those which do not depend on the central 
value of the measurement. However, for cross sections, many uncertainties 
are proportional to the central value (so-called {\it multiplicative} uncertainties). 
This proportionality can be approximated by a linear dependence. In such cases, 
the data combination using Eq.~\ref{eqn:chi2} will introduce a bias 
towards lower cross sections since the measurements with smaller central 
values will have smaller absolute uncertainties. An improved $\chi^2$ can be defined by 
replacing, in Eq.~\ref{eqn:chi2}, $\delta_i \rightarrow \frac{M^{i,{\rm true}}}{M^i}\delta_i$ 
and $\frac{\partial M^i}{\partial \alpha_j}\alpha_j \rightarrow \frac{\partial M^i}
{\partial \alpha_j} \frac{M^{i,{\rm true}}}{M^i} \alpha_j$ which translates the 
multiplicative uncertainties for each measurement to the absolute ones. 
Unlike Eq.~\ref{eqn:chi2}, however, this $\chi^2$ is not a 
simple quadratic function with respect to $\{ M^{i,true} \}$,$\{ \alpha_j \}$. 
Therefore, the minimum is found by an iterative procedure: 
first, Eq.~\ref{eqn:chi2} is used to find an initial approximation 
for $ \{ M^{i,true} \}$, then the uncertainties are re-calculated 
using $\delta_i \rightarrow \frac{M^{i,true}}{M^i}\delta_i$,  
and finally the minimisation is repeated. 
For the HERA data averaging, convergence was observed after two iterations.

\subsection{Data input and treatment}
The data used for the combination consist of the published double differential NC and CC 
cross sections from H1 and ZEUS, taken in the years $1994-2000$, 
and are listed in Tab.~\ref{tab:ncccdatasets}. During this period, HERA operated with 
an electron beam energy, $E_e$, of 27.5 GeV and 
a proton beam energy, $E_p$, of 820 GeV (until 1997) and 920 GeV (from 1998 onwards). 
The measurements span the kinematic region $1.5 < Q^2 < 30000$ ${\rm GeV}^2$ and 
$6.5 \times 10^{-5} < x < 0.65$, and are the most precise data 
published by the H1 and ZEUS collaborations to date\footnote{
Note that there are also data available for $Q^2<1$ ${\rm GeV}^2$, both from shifted vertex 
operation and from ZEUS using a dedicated detector near the beam pipe. These data have not 
been considered here, but will be included in subsequent combined data analyses.}.

The double differential cross section measurements are 
published with their statistical 
and systematic uncertainties. The statistical uncertainties 
are uncorrelated between different data points. 
The systematics are classified into three sub-groups: 
(i) point-to-point uncorrelated systematics, (ii) 
point-to-point correlated (e.g. energy scale calibrations), 
(iii) an overall normalisation uncertainty of various data sets. 
Sources of point-to-point correlated uncertainties are often 
common for NC and CC cross section measurements and sometimes 
can be considered to be correlated for different data sets of 
the same experiment. They are treated as independent between 
H1 and ZEUS, since uncertainties of beam energies are negligible. 
Similarly, the normalisation uncertainties are correlated 
for all cross section measurements by a given experiment 
from a common data-taking period. 

All the NC and CC cross section data from H1 and ZEUS are 
combined in one simultaneous minimisation. Therefore, resulting 
shifts of correlated systematic uncertainties and global 
normalisations propagate coherently to both NC and CC data.

\begin{figure}[Htp]
\vspace{-0.5cm}
\includegraphics[width=8cm,height=9cm]{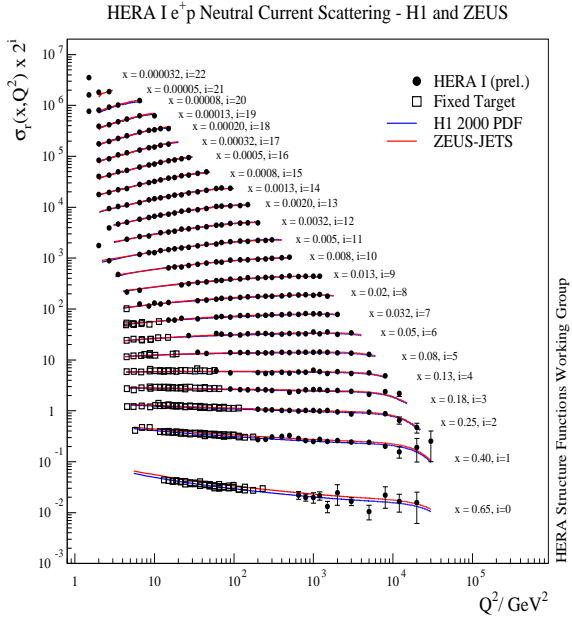}
\vspace{-1.2cm}
\caption{Deep inelastic neutral current $e^+p$ scattering cross section data  
from the HERA I data-taking period as obtained by combining the published H1 and ZEUS measurements. 
The curves are NLO QCD fits as performed by H1 and ZEUS to their own data.}
\label{fig:nce+p}
\end{figure}

\subsubsection{Extrapolation to common $x-Q^2$}
Prior to the combination, the H1 and ZEUS data were transformed 
to a common grid of $(x,Q^2)$ points using a simple interpolation:
\begin{eqnarray}
\sigma_{ep}(x_{grid},Q^2_{grid}) 
= \frac{\sigma^{th}_{ep}(x_{grid},Q^2_{grid})}{\sigma^{th}_{ep}(x,Q^2)} 
                                    \sigma_{ep}(x,Q^2). \nonumber
\end{eqnarray}
The H1 PDF parameterisation~\cite{epj:c30:32} of the double differential 
NC and CC cross sections was used to calculate the theoretical ratios.   
The sensitivity of the data combination 
to the choice of parameterisation was checked using 
the ZEUS PDFs~\cite{epj:c42:1}. The resulting correction 
factors were found to agree to within a few permille for the NC data and to within 
$2\%$ for the CC data (i.e. in both cases, much less than the experimental uncertainties). 
\begin{figure}[Htp]
\vspace{-0.5cm}
\includegraphics[width=8cm,height=8cm]{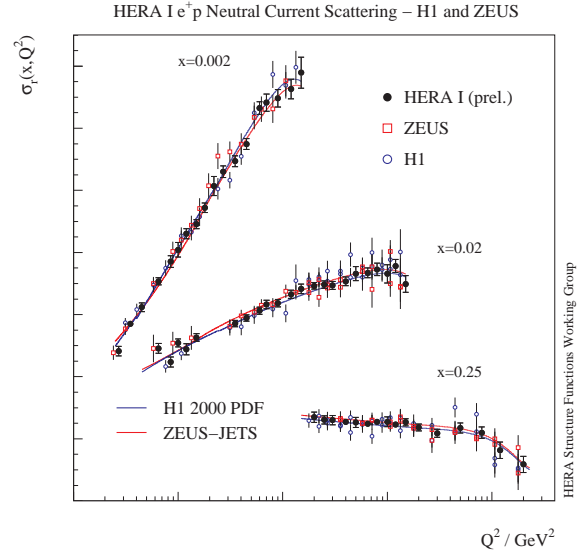}
\vspace{-1.2cm}
\caption{Neutral current $e^+p$ cross section measurements for three 
selected $x$ bins as a function of $Q^2$. The H1 (open points), ZEUS 
(open squares) are compared to the H1 and ZEUS combined data (solid points). 
Measurements from the individual experiments have been shifted for clarity. 
The error bars indicate the total uncertainty. The curves show the predictions 
of NLO QCD fits performed by H1 and ZEUS, on their own data.}
\label{fig:nce+pzoom}
\end{figure}

\subsubsection{Centre-of-mass energy correction}
The data sets listed in Tab.~\ref{tab:ncccdatasets}, 
include samples from both $E_p=820$ GeV and $E_p=920$ GeV running. 
Since the NC and CC DIS cross sections depend weakly on the 
energy, via terms involving the inelasticity $y$, 
a choice must be as to whether to keep the two beam-energy samples separate, 
or to correct to a single, common proton beam energy. 
For the present combination, the latter choice was made such 
that the NC and CC data at $E_p=820$ GeV were transformed 
to $E_p=920$ GeV and then combined with the genuine 
measurements at the higher proton beam energy.

For the CC data , the correction was calculated as: 
\begin{eqnarray}
\sigma_{CC,920}^{ep}(x,Q^2) 
= \frac{\sigma^{th,ep}_{CC,920}(x,Q^2)}{\sigma^{th,ep}_{CC,820}(x,Q^2)} 
                                  \sigma_{CC,820}^{ep}(x,Q^2). \nonumber
\end{eqnarray}
For the NC data, the correction was performed additively:
\begin{eqnarray}
\sigma_{NC,920}^{ep}(x,Q^2) &=&  \sigma_{NC,820}^{ep}(x,Q^2) \nonumber \\ 
&+& \Delta\sigma_{NC,920}^{ep}(x,Q^2,y_{920},y_{820}), \nonumber
\end{eqnarray}
where the correction term is the difference between H1 PDF~\cite{epj:c30:32} 
based predictions of the DIS cross sections, 
with $y_{920}=Q^2/(4xE_e920)$ and $y_{820}=Q^2/(4xE_e820)$. 
The corrections were found to only be sizable at large $y$. To estimate 
the uncertainty on the combined data due to this procedure, 
another average was calculated assuming $F_L=0$ as an extreme 
assumption. The difference between these results and those 
using the standard choice were at the permille level across 
most of the kinematic plane, and reached $\sim5\%$ for only a very 
few points at $y>0.6$. This uncertainty was added in quadrature 
to the combined data.

\subsubsection{Additive vs multiplicative errors}
As mentioned in Sec.~\ref{sec:combinationmethod}, uncertainties 
on cross section measurements may be 
additive or multiplicative. Within the community it is generally agreed  
that normalisation uncertainties are multiplicative. 
However, for the other systematic uncertainties the situation is less clear. 
To assess the sensitivity of the average HERA data set to this issue, 
various different treatments of the systematic uncertainties were considered. 
The extreme assumptions treat all uncertainties as multiplicative, 
or all as additive, apart from the normalisation uncertainties. 
Therefore, an additional systematic uncertainty was estimated, based on the 
difference between these two error treatments. The typical size 
of this uncertainty was $<1\%$ for the low $Q^2$ data, reaching $1-1.5\%$ 
at larger $Q^2$.

\subsubsection{Correlations between experiments}
The H1 and ZEUS collaborations use similar methods to reconstruct 
the event kinematics, employ similar techniques for the detector 
calibration, use common Monte Carlo simulation models for the 
hadronic final state simulation as well as for photoproduction 
background subtraction. This similarity of approaches and techniques 
may lead to correlations between H1 and ZEUS measurements. 

A detailed investigation has shown that the results of the combination are rather 
insensitive to the assumptions on correlations between the two 
experiments. The largest effect on the average derives from differing assumptions 
on the photoproduction background (a $1-2\%$ change at $y>0.6$ for 
low $Q^2<20$ ${GeV}^2$) and on the hadronic energy calibration 
($1\%$ at low $y<0.02$). For these sources the measurements rely 
more on the simulation of the hadronic final state which is similar 
for the two experiments. These variations are therefore introduced 
as additional point-to-point correlated systematic sources of 
uncertainty on the averaged cross sections.

\subsection{Results}

In the minimisation procedure, 1153 individual NC and CC measurements were averaged 
to 584 unique points. This yielded a good quality of fit with the $\chi^2/dof$ = $510/599$. 
The distribution of pulls did not show any significant tension across the kinematic plane. 
A total of 43 sources of correlated systematic uncertainty were considered in this analysis. 
In the combined data, almost all systematic uncertainties were reduced, with the most significant 
reduction observed for the H1 backward calorimeter energy scale (by a factor of 3), 
and the ZEUS uncertainty in modelling the forward hadronic energy flow (by a factor of 4). 
The errors which result from the combination process (centre-of-mass energy correction, 
additive versus multiplicative error treatment, correlations in the background subtraction 
and on hadronic energy calibration) were introduced as additional point-to-point correlated 
systematic sources of uncertainty on the average cross section.

Figure~\ref{fig:nce+p} shows the complete set of combined $e^+ p$ NC data, 
spanning the entire $x$ and $Q^2$ range. The scaling violations are clearly 
visible at both low and high $x$. The data are compared to two previously published 
QCD fits performed by the H1 and ZEUS collaborations on their own data. 
As expected, the fits provide an excellent description of the HERA combined data.

Figure~\ref{fig:nce+pzoom} shows a close up, with a linear $y$ scale, of three selected $x$ bins, 
as a function of $Q^2$. The combined data are compared to the individual measurements from 
H1 and ZEUS (shifted for clarity). At low $Q^2$, where the data are limited 
by systematic uncertainties, the improvement in the total uncertainty is visible. 
At higher $Q^2$, the combined data have significantly reduced uncertainties, 
and exhibit far smaller fluctuations, which is driven by the increase in 
statistical accuracy, which dominates the measurement.

The combined NC and CC HERA data have subsequently been included in a NLO QCD 
fit to extract the proton PDFs. Details of this analysis are described 
in the next section.

\section{QCD analysis of the combined data}
\label{sec:qcdanalysis}
Previously, the H1 and ZEUS collaborations have both used their own data 
in NLO QCD fits~\cite{epj:c30:32,epj:c42:1}. These data sets have very small 
statistical uncertainties, so that the contribution 
from systematics becomes dominant and consideration of point-to-point correlated 
systematic uncertainties is essential. The ZEUS analysis takes account of correlated experimental 
systematics by the Offset method (see e.g.~\cite{epj:c42:1}), 
while H1 uses the Hessian method~\cite{jphys:g28:2669}. 
In an attempt to improve the determination of the PDFs from HERA data, 
the combined H1 and ZEUS measurements 
have been used as the sole 
input for a new, NLO QCD analysis.
In Sec.~\ref{sec:qcdfit-analysis}, the QCD analysis, model assumptions and treatment of 
uncertainties are discussed. In Sec.~\ref{sec:qcdfit-results}, the results are presented.

\subsection{NLO QCD analysis}
\label{sec:qcdfit-analysis}
For the QCD analysis presented here 
the predictions for the structure functions were obtained by solving the 
DGLAP evolution equations at NLO, in the $\overline{\rm MS}$ scheme, with the 
renormalisation and factorisation scales taken to be $Q^2$. The DGLAP equations 
yield the PDFs at all values of $Q^2$ provided they are input as functions of $x$ at some input scale $Q_0^2$. 
For this analysis, the input scale was chosen to be $Q_0^2=4$ ${\rm GeV}^2$. 
The resulting PDFs were then convoluted with NLO coefficient functions 
to give the structure functions which enter into the expression for the cross sections. 
The choice of the heavy quark masses were $m_c=1.4$ GeV and $m_b=4.75$ GeV. For this 
preliminary analysis, the heavy quark coefficient functions have been calculated 
in the zero-mass-variable-flavour-number scheme. The strong coupling constant 
was fixed at $\alpha_s(M_Z)=0.1176$~\cite{jphys:g33:1}.

The fit was performed at leading twist. Since the HERA data have a minimum 
invariant mass squared of the hadronic system, $W^2$, of $W^2_{min}=300$ ${\rm GeV}^2$ 
and a maximum $x$ of $x_{max}=0.65$, they are in a kinematic region 
where there is no sensitivity to target mass and large-$x$ higher twist contributions. 
However, a minimum $Q^2$ cut of $Q^2_{min}=3.5$ ${\rm GeV}^2$  has been imposed on 
the data included in the fit, in order to remain in the kinematic regime 
where perturbative QCD should be applicable.

\subsubsection{Choice of parameterisation}
\label{sec:parameterisation}
The PDFs were parameterised using the form:
\begin{eqnarray}
xf(x)=Ax^B(1-x)^C(1+Dx+Ex^2+Fx^3), \nonumber
\end{eqnarray}
and the number of parameters was chosen such that $D$, $E$ and $F$ 
were only varied if this resulted in a significant improvement to the $\chi^2$. 
Otherwise, these parameters were set to zero.

For the present analysis\footnote{Note that the choice of 
parameterisation described in Sec.~\ref{sec:parameterisation} 
has been inspired by both the H1- and the ZEUS-style parameterisations. 
It attempts to combine the best features of both,  
in that it has fewer assumptions concerning $x(\bar{d}-\bar{u})$ than 
the ZEUS-style and less model dependence than the H1-style since it 
does not assume equality of all $B$-parameters.}, the following PDFs 
were parameterised: $xu_v$, $xd_v$, 
$xg$, $x\overline{U}=x(\bar{u}+\bar{c})$ and  $x\overline{D}=x(\bar{d}+\bar{s})$ such that:
\begin{eqnarray}
xu_v(x) &=& A_{u_v}x^{B_{u_v}}(1-x)^{C_{u_v}}(1+D_{u_v}x+E_{u_v}x^2)  \nonumber \\
xd_v(x) &=& A_{d_v}x^{B_{d_v}}(1-x)^{C_{d_v}} \nonumber \\
x\overline{U}(x) &=& A_{\overline{U}}x^{B_{\overline{U}}}(1-x)^{C_{\overline{U}}} \nonumber \\
x\overline{D}(x) &=& A_{\overline{D}}x^{B_{\overline{D}}}(1-x)^{C_{\overline{D}}} \nonumber \\
xg(x) &=& A_{g}x^{B_{g}}(1-x)^{C_{g}}. \nonumber 
\end{eqnarray}
The normalisation parameters, $A_{u_v}$ and  $A_{d_v}$ were constrained to 
satisfy the number sum rules and $A_g$ was constrained to satisfy the 
momentum sum rule. The $B$ parameters, $B_{u_v}$ and $B_{d_v}$ were set equal, $B_{u_v}= B_{d_v}$, 
as were the $B_{\overline{U}}$ and $B_{\overline{D}}$,  $B_{\overline{U}}= B_{\overline{D}}$, 
such that a single $B$-parameter was used for the valence quarks and another,  
different single $B$-parameter was used for the sea distributions. A further constraint 
was provided by assuming that the strange and charm quark distributions can 
be expressed as $x$ dependent fractions, $f_s=0.33$ and $f_c=0.15$, of the $d$ and $u$ type 
sea, such that $A(\overline{U})=A(\overline{D})(1-f_s)/(1-f_c)$. The value of 
$f_s=0.33$ was chosen to be consistent with determinations of this fraction from 
neutrino induced dimuon production. The charm fraction was set to be consistent 
with dynamic generation of charm from the start point of $Q_0^2=m_c^2$ 
in a zero-mass-variable-flavour-number scheme. In total, there were 11 free parameters in the fit. 
\begin{center}
\begin{table}[Htp]
\begin{tabular}{|c|c|c|c|}
\hline
{\small model variation} & central value   & lower   & upper  \\ \hline
{\small $m_c$} & 1.4  & 1.35 & 1.5 \\
{\small $m_b$} & 4.75 & 4.3  & 5.0 \\  
{\small $Q^2_{min}$} & 3.5 & 2.5  & 5.0 \\  
{\small $Q_0^2$} & 4.0 & 2.0  & 6.0 \\  
{\small $f_s$} & 0.33 & 0.25  & 0.40 \\  
{\small $f_c$} & 0.15 & 0.12  & 0.18 \\  \hline
\end{tabular}
\vspace{0.3cm}
\caption{Central values of input parameters and cuts, and the variations considered to evaluate model uncertainties.
}
\label{tab:model}
\end{table}
\end{center}

\subsubsection{Treatment of experimental data}
The HERA combined $e^\pm p$ NC and CC inclusive data, described in 
Sec.~\ref{sec:combineddata}, have been used as the experimental input to the 
present QCD analysis\footnote{Note that the version of the combined data used 
for the QCD analysis is slightly different to that described in 
Sec.~\ref{sec:combineddata} in that only data with $y<0.35$ were 
transformed in $E_p$.}. The consistency of the combined HERA data set, 
as well as its small systematic uncertainties, allow the experimental 
uncertainties on the PDFs to be calculated using a $\chi^2$ tolerance 
of $\Delta\chi^2 = 1$. This is compared to previous global fits where 
increased tolerances of  $\Delta\chi^2 = 50-100$ have been used 
in order to account for data inconsistencies.
\begin{figure*}[Htp]
\includegraphics[width=7.8cm,height=9.cm]{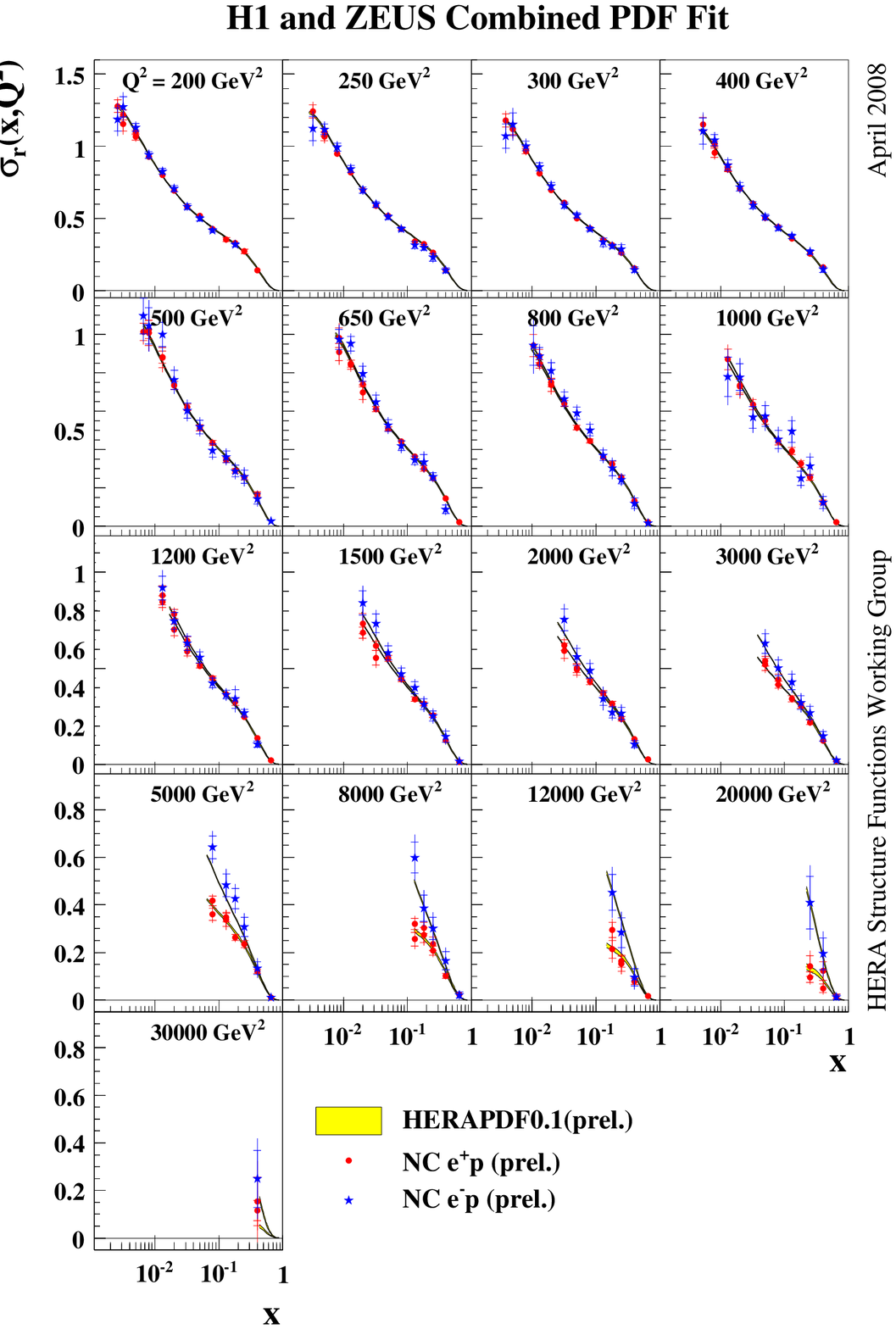}
\includegraphics[width=7.8cm,height=9.cm]{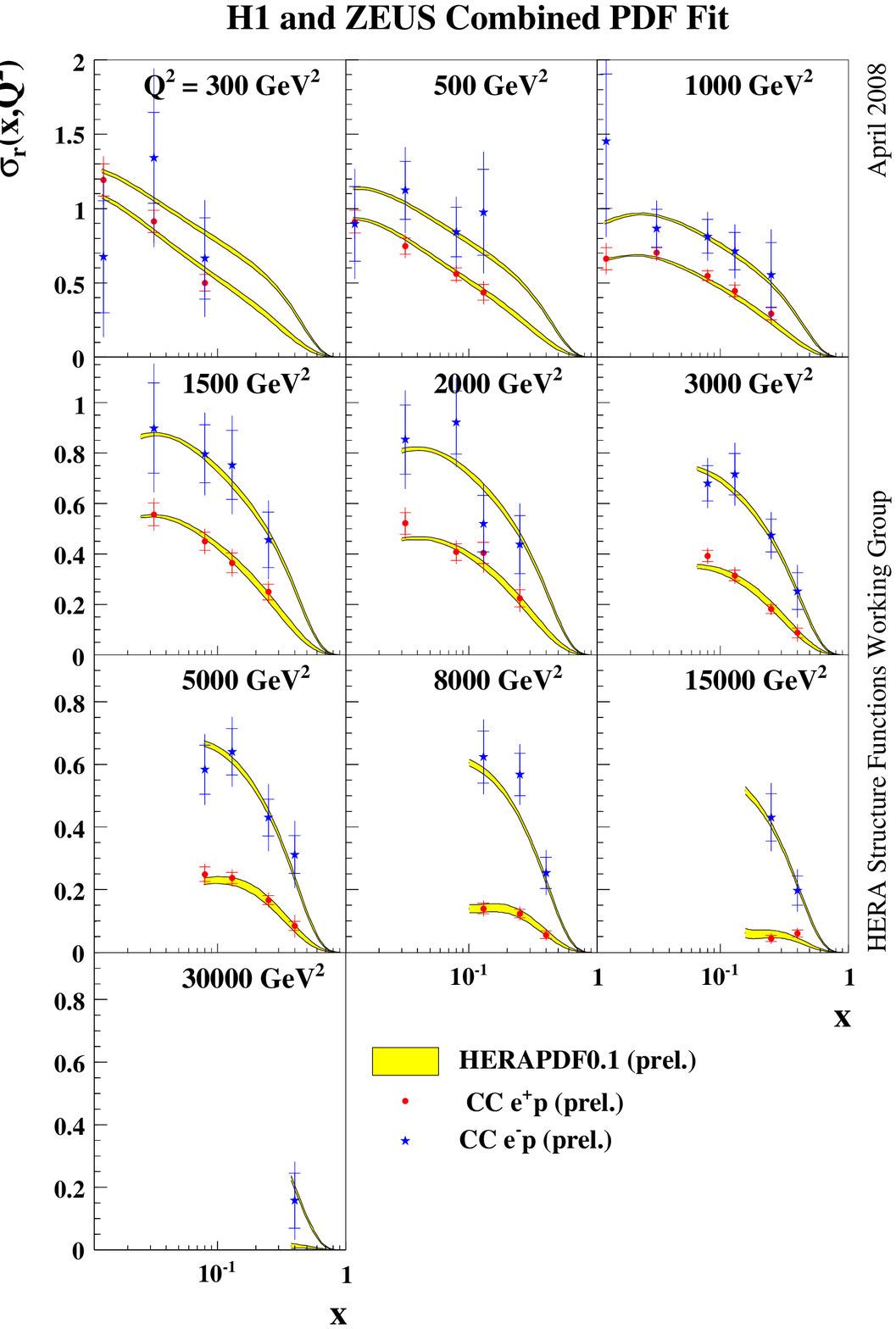}
\vspace{-1.2cm}
\caption{HERA combined NC (left) and CC (right) reduced cross section data. 
The predictions of the HERAPDF0.1 fit are superimposed. 
The uncertainty bands illustrated derive from both experimental and model sources.
}
\label{fig:nccc-herafit}
\end{figure*}
\begin{figure*}[Htp]
\vspace{0.2cm}
\hspace{0.3cm}\includegraphics[width=7.8cm,height=7.8cm]{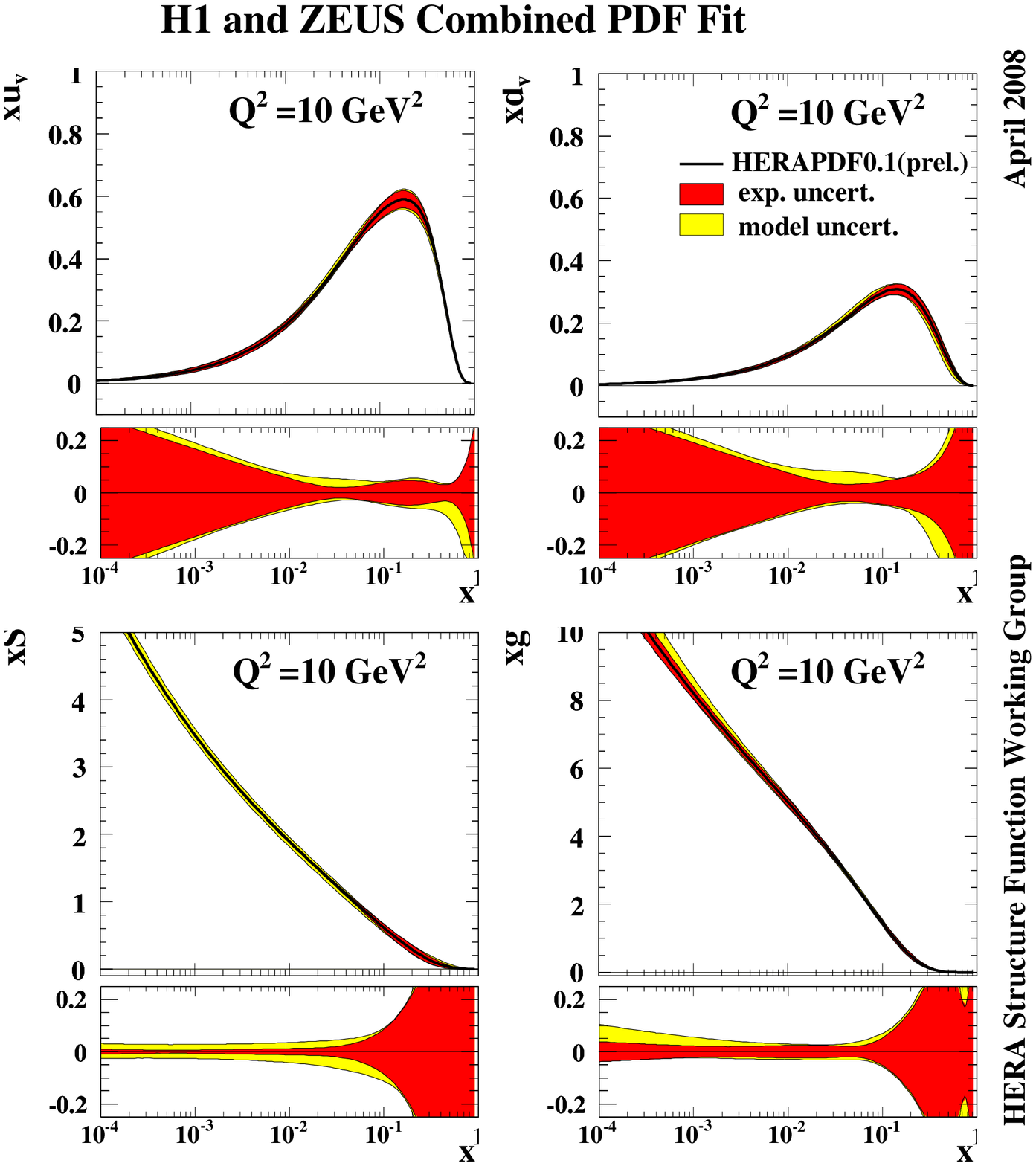}
\includegraphics[width=7.8cm,height=7.8cm]{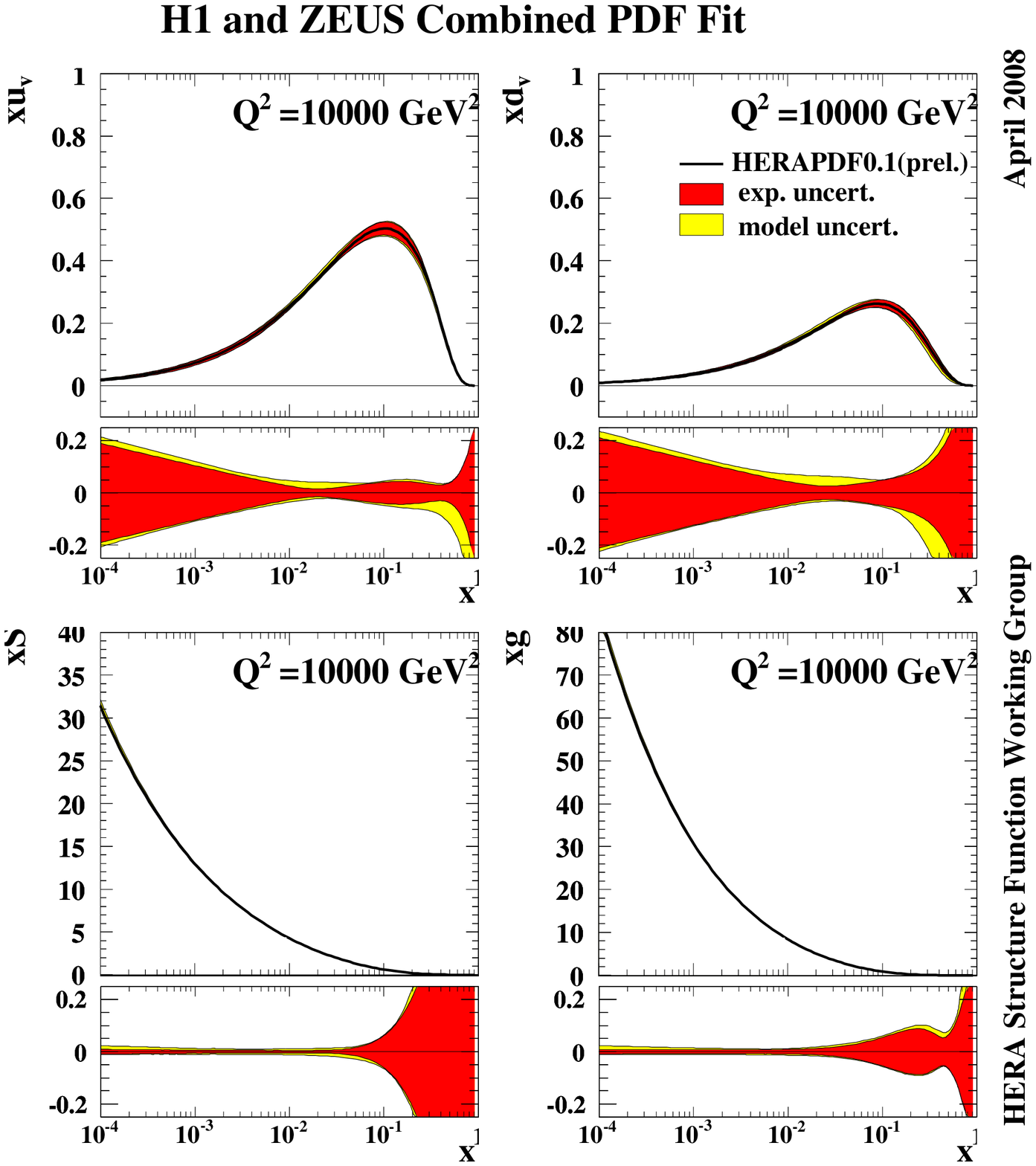}
\vspace{-1.cm}
\caption{The HERA PDFs $xu_v$, $xd_v$, $xS$ (total sea-quark distribution) and $xg$ at 
(left) $Q^2=10$ ${\rm GeV}^2$ and (right) $Q^2=10000$ ${\rm GeV}^2$. 
Fractional uncertainty bands are shown beneath each PDF.  
The experimental and model uncertainties are shown separately. 
}
\label{fig:herapdf}
\end{figure*}

For the present QCD fit, the role of correlated systematics is no longer crucial 
since the uncertainties are relatively small. In particular, this means that 
similar results are obtained whether the correlated systematic uncertainties 
are treated using the Offset or Hessian method in the QCD fit, or by 
simply combining statistical and systematic uncertainties in quadrature. 
For the central fit the choice was made to combine the 43 systematic 
uncertainties (resulting from the separate H1 and ZEUS data sets)  
in quadrature and to Offset the 4 sources of uncertainty which arise 
from the combination procedure. This was found to result in the most conservative 
uncertainty estimate on the extracted PDFs. 
\begin{figure*}[Htp]
\hspace{0.7cm}\includegraphics[width=6.8cm,height=7.2cm]{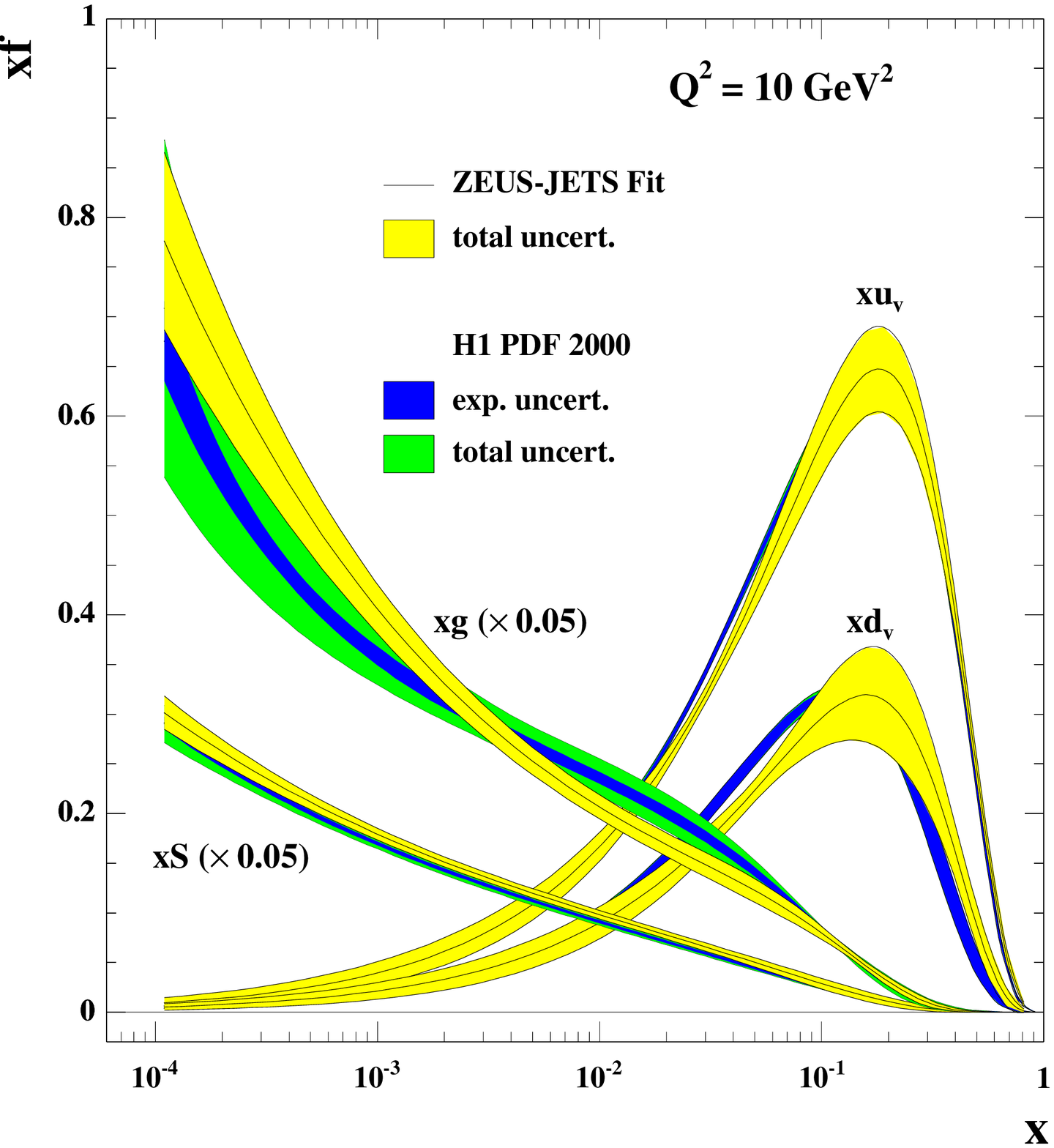}
\includegraphics[width=7.5cm]{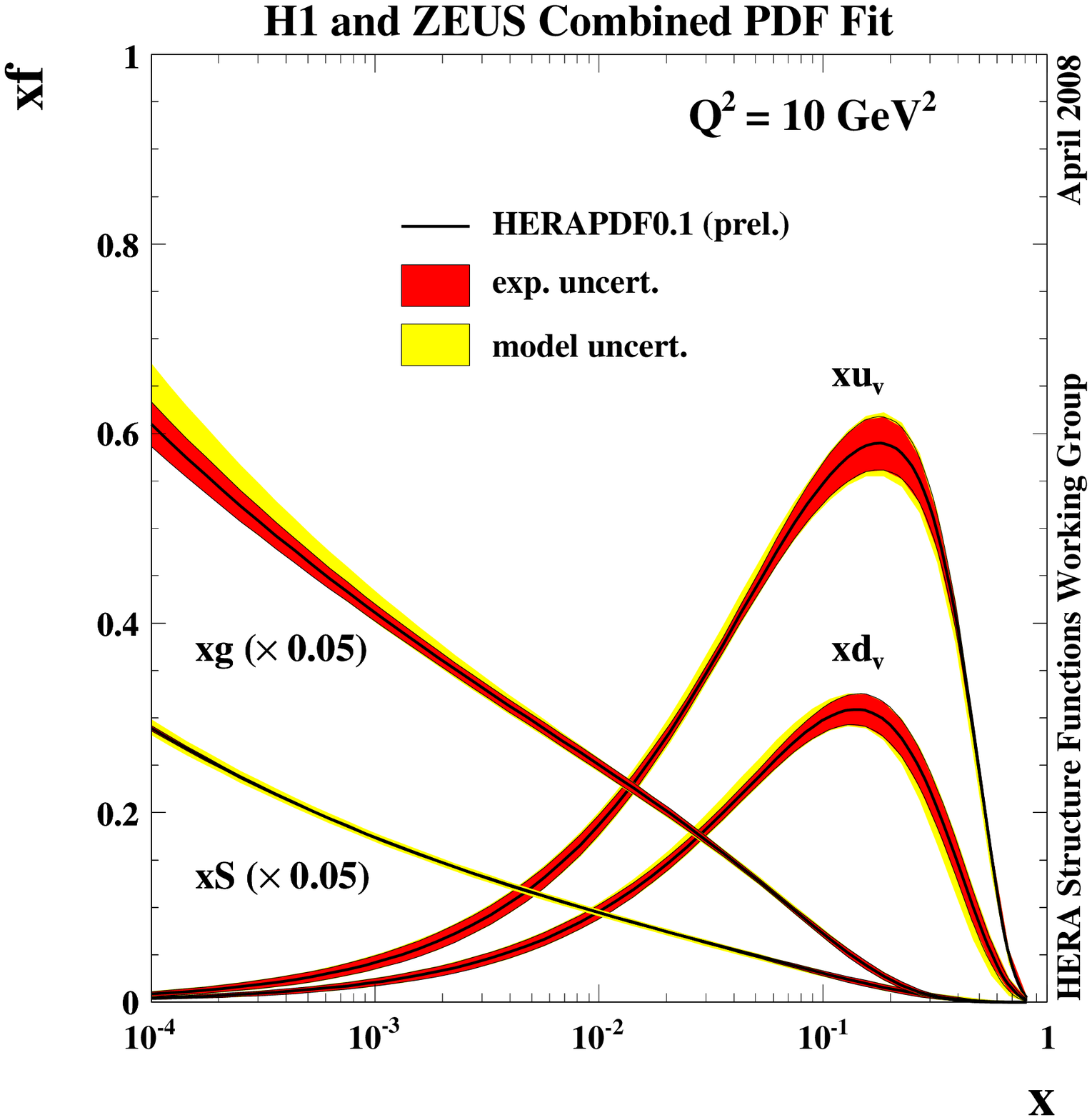}
\vspace{-0.8cm}
\caption{Left: The H1 and ZEUS PDFs obtained from separate QCD analyses, each to their own data.
Right: The HERAPDF0.1 PDFs obtained from the analysis of the combined data set, as described here.
}
\label{fig:herapdfsummary}
\end{figure*}
\begin{figure*}[Htp]
\includegraphics[width=7.5cm]{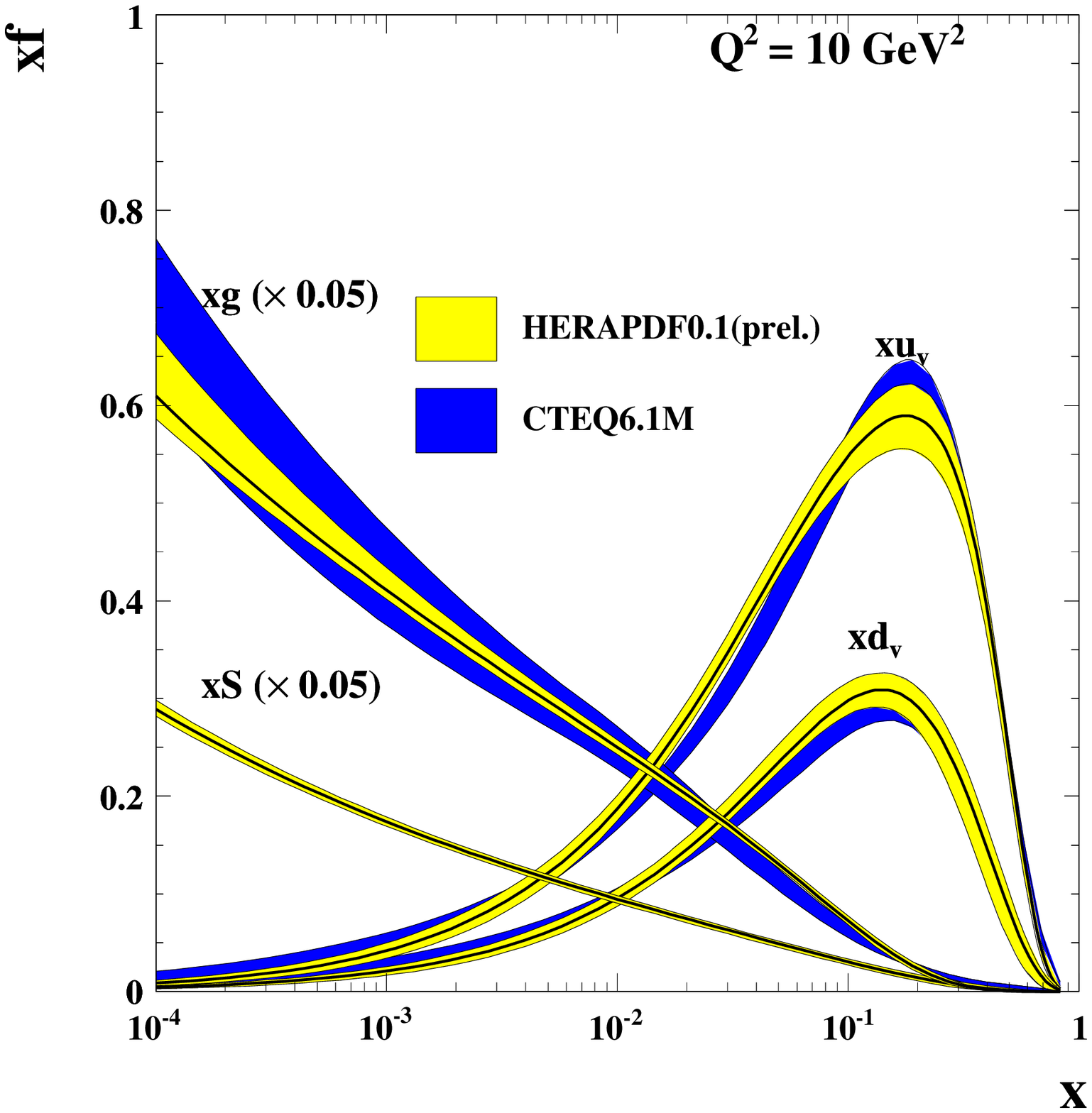}
\includegraphics[width=7.5cm]{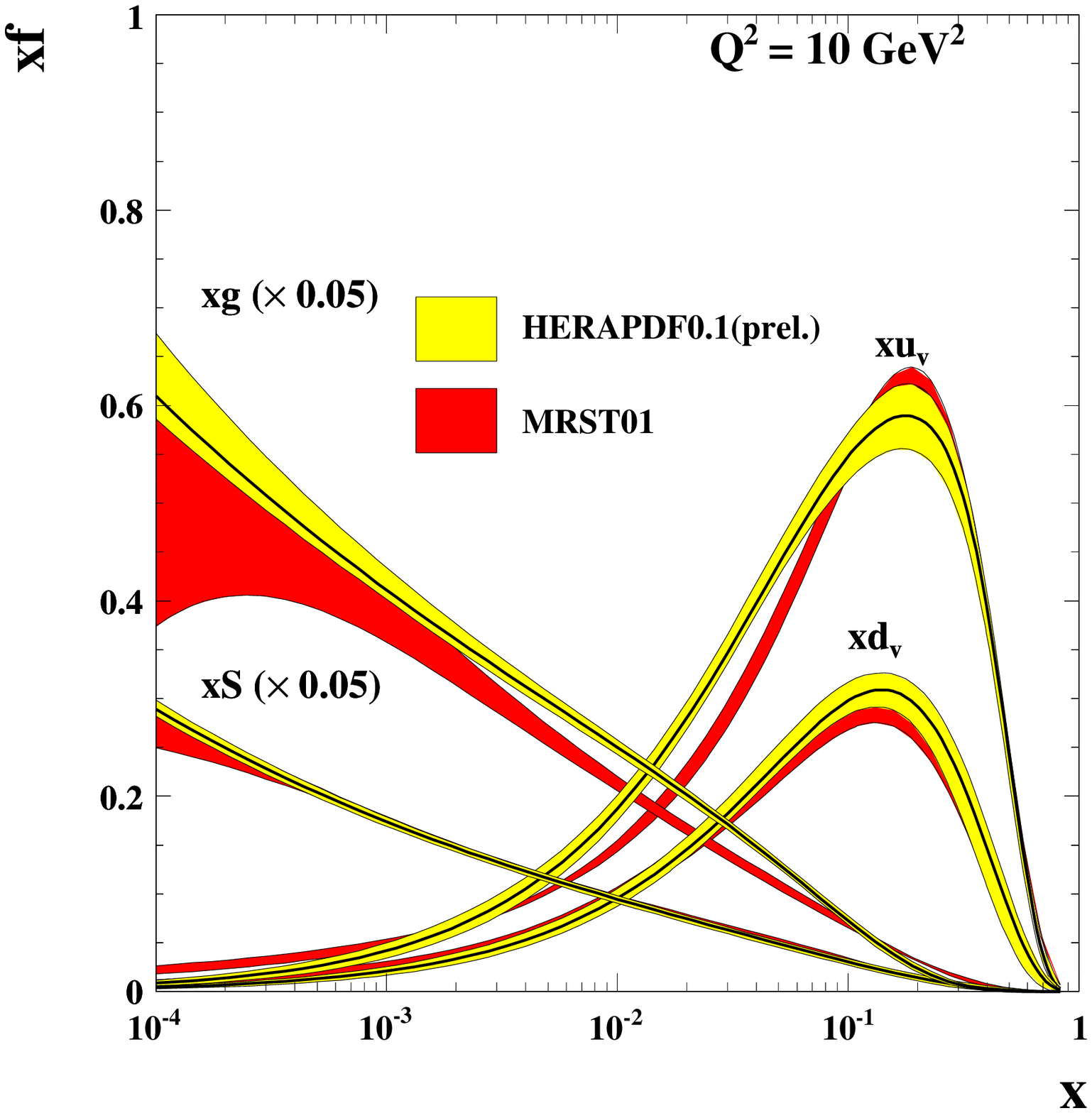}
\vspace{-0.8cm}
\caption{The HERAPDF0.1 PDFs compared to the CTEQ6.1M and MRST01 fits at $Q^2=10$ ${\rm GeV}^2$.}
\label{fig:globalpdfs}
\end{figure*}

\subsubsection{Model uncertainties}
Despite the conservative procedure adopted, as mentioned above,  
the experimental uncertainties on the resulting PDFs were still 
found to be impressively small. Therefore, a thorough consideration 
of further uncertainties due to model assumptions was necessary. 
For the present analysis, six sources of model uncertainty were considered, 
as listed in Tab.~\ref{tab:model}. 
The parameters were varied up and down from the central value, 
and the differences in the resulting PDFs from the central fit 
were added in quadrature to the total experimental PDF uncertainty. 

Further cross checks were also performed. In particular, the dependence on the 
choice of parameterisation was investigated by repeating the fit using the 
H1-style \cite{epj:c30:32} and ZEUS-style \cite{epj:c42:1} parameterisations.
All fits gave acceptable $\chi^2$ and were found to be consistent with each other. 
Note, however, that the parameterisation chosen for the central fit 
(described in Sec.~\ref{sec:qcdfit-analysis}) gave the best $\chi^2$ as well as 
the most conservative experimental uncertainties. 

\subsection{Results}
\label{sec:qcdfit-results}
The NLO QCD analysis described above has been named the HERAPDF0.1 fit. 
Figure~\ref{fig:nccc-herafit} shows the results of the HERA fit superimposed 
on the combined high-$Q^2$ NC and CC $e^\pm p$ data sets, showing 
the excellent description of the data by the HERA PDFs. 

Figure~\ref{fig:herapdf} shows the HERAPDF0.1 PDFs $xu_v$, $xd_v$, 
$xS$ (total sea) and $xg$ 
as a function of $x$ at $Q^2=10$ ${\rm GeV}^2$ and $Q^2=10000$ ${\rm GeV}^2$. 
Fractional uncertainty bands are shown beneath each plot, with the experimental 
and model uncertainties being shown separately. The variation of the 
strange fraction, $f_s$, dominates the model uncertainty on the sea, while 
variations of $Q_0^2$ and $Q^2_{min}$ dominate those on the gluon and 
valence quarks. Comparison of the results for $Q^2=10$ ${\rm GeV}^2$ and 
$Q^2=10000$ ${\rm GeV}^2$ shows that, as $Q^2$ increases, 
the PDF uncertainties become impressively small. 

The summary plots shown in Fig.~\ref{fig:herapdfsummary} illustrate that the total uncertainty 
of the PDFs obtained from the HERA combined data set is much reduced compared to the those of the  
PDFs extracted from separate analyses of the H1 and ZEUS data sets. 
The dramatic improvement is a result of the data combination. 
Figure~\ref{fig:globalpdfs} compares the HERAPDF 0.1 PDFs to those of the CTEQ6.1~\cite{jhep:0207:012}
and MRST01~\cite{epj:c23:73} global fits\footnote{Note that the HERAPDF0.1 uncertainty 
band represents a $68\%$ confidence level, while the global fits show a $90\%$ 
confidence level band.}. The results indicate that the precision of the 
HERAPDF0.1 PDFs for the low $x$ sea and gluon is impressive.

\section{Summary}
A new, model-independent method of combining cross section measurements 
has been presented, in which a coherent treatment of systematics 
results in a substantial reduction in the overall uncertainties. 
The method has been demonstrated on the complete set of published HERA I 
NC and CC inclusive DIS data.  

The combined HERA data have subsequently been included in a new NLO QCD analysis. 
The consistent treatment of systematic uncertainties in the combined data set 
ensures that the experimental uncertainties on the PDFs can be calculated 
without need for an increased $\chi^2$ tolerance. This results in PDFs with 
greatly reduced experimental uncertainties compared to separate analyses of data from the 
H1 and ZEUS experiments. Model uncertainties have also been carefully considered. 
The resulting PDFs (called HERAPDF0.1) have impressive precision. 
They are now publicly available in LHAPDF~\cite{ref:lhapdf} (v5.6 onwards).

High statistics data from the HERA II running period are being analysed 
by the two collaborations and will be included in subsequent analyses 
devoted to precision determinations of the proton parton densities.

\section{Acknowledgements}
I would like to thank the organisers of Ringberg 2008 
for a very productive and enjoyable workshop, the members of the 
HERA Combined Structure Function Working Group for useful 
information and comments, and the UK Science and Technology 
Facilities Council for support.

\end{document}